\documentclass[aps,preprintnumbers,showpacs,showkeys]{revtex4}
\usepackage{amssymb,amsmath,amsfonts,amsthm,graphicx}
\usepackage[T1]{fontenc}
\usepackage[utf8]{inputenc}
\usepackage{epsfig}
\graphicspath{{./Figures/}} 
\usepackage{color}
\usepackage{multirow}
\usepackage{array}
\usepackage{picture}

\newcommand{\beq}{\begin{equation}}
\newcommand{\eeq}{\end{equation}}
\newcommand{\bea}{\begin{eqnarray}}
\newcommand{\eea}{\end{eqnarray}}

\newcommand{\bc}{\begin{center}}
\newcommand{\ec}{\end{center}}

\newcommand{\ds}{\displaystyle}

\newcommand*\xbar[1]{%
  \hbox{%
    \vbox{%
      \hrule height 0.5pt 
      \kern0.5ex
      \hbox{%
        \kern-0.1em
        \ensuremath{#1}%
        \kern-0.1em
      }%
    }%
  }%
}

\usepackage{amsmath}
\newcommand{\lr}[1]{ \left( #1 \right) }
\newcommand{\lrs}[1]{ \left[ #1 \right] }
\newcommand{\lrc}[1]{ \left\{ #1 \right\} }

\newcommand{\Tr}{{\mathrm{Tr}\,}}
\newcommand{\ReP}{{\mathrm{Re}\,}}

\usepackage[dvipsnames]{xcolor}
\colorlet{LightRubineRed}{RubineRed!70!}
\colorlet{Mycolor1}{green!10!orange!90!}
\definecolor{Mycolor2}{HTML}{00F9DE}

\begin{document}
\title{Study of two color QCD on large lattices}
\author{A. Begun}
\affiliation{Pacific Quantum Center, Far Eastern Federal University, 690950 Vladivostok, Russia}
\author{V.~G.~Bornyakov}
\affiliation{Institute for High Energy Physics NRC Kurchatov Institute,
142281 Protvino, Russia, \\
Institute of Theoretical and Experimental Physics NRC Kurchatov Institute, 117218 Moscow, Russia
}
\author{V.~A. Goy}
\affiliation{Pacific Quantum Center, Far Eastern Federal University, 690950 Vladivostok, Russia,\\
Institute of Theoretical and Experimental Physics NRC Kurchatov Institute, 117218 Moscow, Russia
}
\author{A. Nakamura}
\affiliation{Pacific Quantum Center, Far Eastern Federal University, 690950 Vladivostok, Russia, \\
Research Center for Nuclear Physics, Osaka University
10-1 Mihogaoka, Ibaraki, Osaka, 567-0047, \\
RIKEN,  Nishina Center, Quantum Hadron Physics Lab.
}
\author{R.~N.~Rogalyov}
\affiliation{Institute for High Energy Physics NRC Kurchatov Institute,
142281 Protvino, Russia}

\begin{abstract}
We study two colors lattice QCD (QC$_2$D)  with two flavors of staggered fermions on $40^4$ and $32^4$ lattices with lattice spacing $a =0.048$~fm 
in the wide range of the quark chemical potential $\mu_q$. 
Our focus is on the confinement-deconfinement transition in this theory.
Thus we compute the string tension from the Wilson loops and the static quark free energy from the Polyakov loops.
We find that the deconfinement transition found earlier in the range $\mu_q \approx 800 - 1000$ MeV is shifted to higher values.
This shift is attributed to decreasing of the lattice spacing used in our simulations in comparison with the earlier study.  
\end{abstract}

\maketitle
\section{Introduction}

The lattice QC$_2$D at nonzero quark chemical potential was studied quite intensively,
see, e.g. \cite{Nakamura:1984uz,Hands:1999md,Kogut:2001if,Kogut:2002cm,Muroya:2002ry,Hands:2006ve,Cotter:2012mb,Boz:2013rca,Braguta:2016cpw,
Holicki:2017psk,Bornyakov:2017txe,Boz:2018crd,Astrakhantsev:2018uzd,Boz:2019enj,Iida:2019rah,Wilhelm:2019fvp,Bornyakov:2020kyz,Iida:2020emi,
Astrakhantsev:2020tdl,Khunjua:2020xws,Kojo:2021knn} 
and references therein. 
Rather high interest to this theory as well as to other QCD-like theories is due to the similarity of their properties 
in some parts of the phase diagram to properties of QCD. Furthermore, such studies provide a laboratory to check 
the methods and approaches which can be also applied to QCD.  

Here we study the deconfinement transition in QC$_2$D. This transition was studied recently in  Ref.~\cite{Bornyakov:2017txe} where it was found in the range $\mu_q \sim 800 - 1000$~MeV. It was concluded in Ref.~\cite{Bornyakov:2017txe} that the obtained result corresponded to zero temperature. In earlier studies \cite{Cotter:2012mb} of this issue  it was shown that the transition position depended on the temperature $T=1/aN_t$. 
This study found that the deconfinement transition position increased from  $\mu_q \approx 500$~MeV up to 800~MeV
when temperature varied from $T \approx 150$~MeV down to $\approx 50$~MeV. It should be noted that the study of Ref.~\cite{Cotter:2012mb} was made at large lattice spacing $a \geq 0.15 $~fm,
while in Ref.~\cite{Bornyakov:2017txe} the lattice spacing $a=0.044$~fm was used. The goal of our study presented here is to clarify if the position of the deconfinement transition changes substantially with temperature even at rather small temperatures using lattices with small lattice spacing. 

It should be noted that we are working on a symmetric lattices which at zero quark density are usually used to study QCD at zero temperature.  As was explained before in \cite{Boz:2019enj,Kojo:2021knn} at large quark density $1/aN_t$ should be considered as temperature even on a symmetric lattices. 

Furthermore, we present in the Appendix our arguments explaining why at large chemical potentials
even at small $T$ the results differ substantially from $T=0$.
Say, at $\mu_q>1$~GeV the physics at $T \approx 100$~MeV or higher is different from physics at zero temperature.

\section{Lattice setup}
We carry out our study on  $40^4$  and $32^4$ lattices for a set of the quark chemical potentials $\mu_q$ in the range $a\mu_q \in (0, 0.5)$. These are the largest lattices (in terms of the number of lattice sites) used so far in the studies of 
lattice QC$_2$D. 
The tree level improved Symanzik gauge action~\cite{Weisz:1982zw} and the improved staggered fermion action with a diquark source 
term~\cite{Hands:1999md} were used in simulations. The explicit expression for the lattice action is as follows
\begin{equation}
  S_{QC_{2}D}=S_G+S_{stag}\,,
\end{equation}
where
\begin{eqnarray}
  S_G & = & \frac{\beta}{2} \lr{c_0 \underset{plaq}{\sum}\, \ReP\Tr\lr{1-U_{plaq}} + c_1 \underset{rt}{\sum}\, \ReP\Tr\lr{1-U_{rt}}}\,, 
\end{eqnarray}

\begin{eqnarray}
  S_{stag} & = & \underset{x}{\sum} \bar{\psi}_{x}
    \lrs{\underset{\mu}{\sum}\frac{\eta_{x,\mu}}{2}\lrc{
    U_{x,\mu} e^{\delta_{\mu,0}\mu_q a} \psi_{x+\hat\mu} -
    U_{x-\hat\mu,\mu}^{\dagger}  e^{-\delta_{\mu,0}\mu_q a} \psi_{x-\hat\mu}}
    + m a\; \psi_{x}}+\\
  \nonumber
  & + & \underset{x}{\sum}\frac{1}{2}\lambda \lrs{\psi_x^T \sigma_2 \psi_x + \bar{\psi}_x \sigma_2 \bar{\psi}_x^T}\,,
\end{eqnarray}
where $c_0$, $c_1$ -- parameters of improved lattice gauge action, $\beta$ -- inverse coupling constant, $U_{x,\mu}$ -- $SU(2)$ link variable, 
 $S_{stag}$ has implicit summation over the flavor index, 
$\eta_{x,\mu}$ -- staggered sign function~\cite{GattringerBook}. In fact we are using improved  staggered quark Dirac operator changing $U_{x,\mu}$  to stout smeared variables as described in Ref.~\cite{PhysRevD.69.054501}.

The lattice configurations were generated at $\beta=1.75$ and quark mass in lattice units $am_q=0.0075$. We used the diquark source 
term coupling $\lambda=0.00075$ which was much smaller than $am_q$.
We do not expect essential change of our results from extrapolation to $\lambda=0$ limit. We found for the lattice spacing
$r_0/a = 9.8(2)$, where $r_0$ is the Sommer parameter \cite{Sommer:1993ce}. To introduce the physical units we chose to use the value  
$r_0 = 0.468(4)$~fm \cite{Bazavov:2011nk}. Then we get $a = 0.048(1)$~fm and for the lattice size in physical units $L_1=1.92$~fm for $40^4$ lattice and $L_2=1.54$~fm for
$32^4$ lattice. Respective temperature values are $T_1=103$~MeV and $T_2=128$~MeV. For the pion mass we found $r_0 m_\pi = 1.62(10)$ or  $m_\pi = 680(40)$~MeV. Later we will also use results of Ref.~\cite{Bornyakov:2017txe} obtained on
lattices with physical size $L_3=1.4$~fm ($T_3=140$~MeV) and pion mass $m_\pi = 740(40)$~MeV.

\section{Confinement-deconfinement transition in $ T - \mu_q$ plane.}

It is known that the Wilson loop has a tiny overlap with the broken string  state \cite{Bolder:2000un}. Thus it can be used to compute the string tension
$\sigma$
even in a theory with dynamical quarks when the respective state is the ground state only for distances up to the string breaking 
distance $r_{br}$. We follow this strategy to determine  $\mu_q$ dependence of $\sigma$. We measure the Wilson loops  after one iteration of the HYP~\cite{Hasenfratz:2001hp}
procedure for the links in direction $\mu=4$ and 100 APE smearing sweeps~\cite{Albanese:1987ds} for links in all spatial directions. The HYP procedure allows to decrease
substantially the static source self-energy at the cost of making incorrect the static potential $V(r)$ dependence on $r$ for $r < 3a$. It is worth mentioning the technical difficulty in computation of the static potential at nonzero $\mu_q$. With increasing $\mu_q$ the gap between the ground state and the excited state decreases and it becomes more and more difficult to extract the ground state value for the large distance $r$.

\begin{figure}[tbh]
\begin{center}
\includegraphics[width=6.5cm, angle=270]{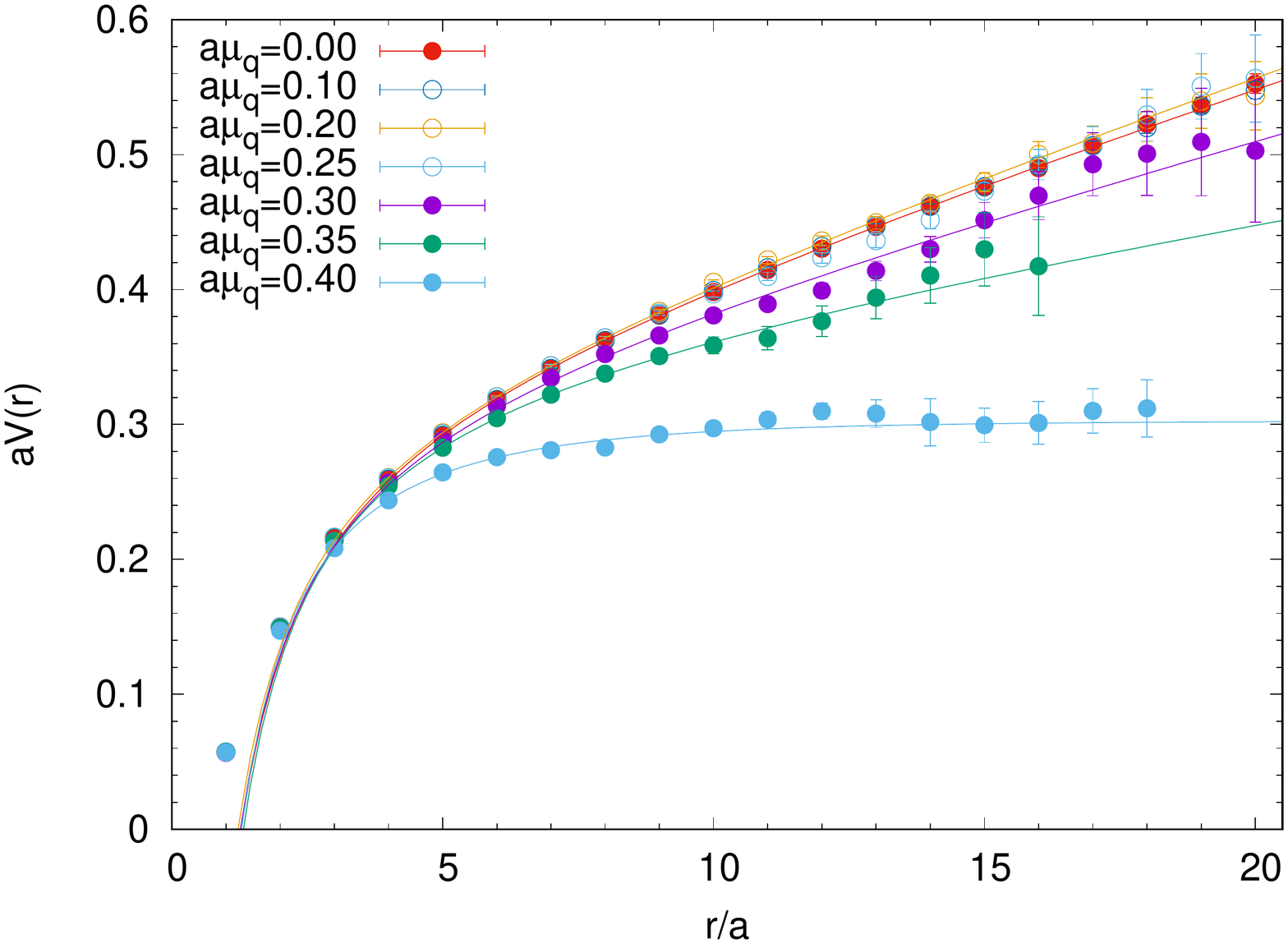} \hspace{-1.7cm}
\includegraphics[width=6.5cm, angle=270]{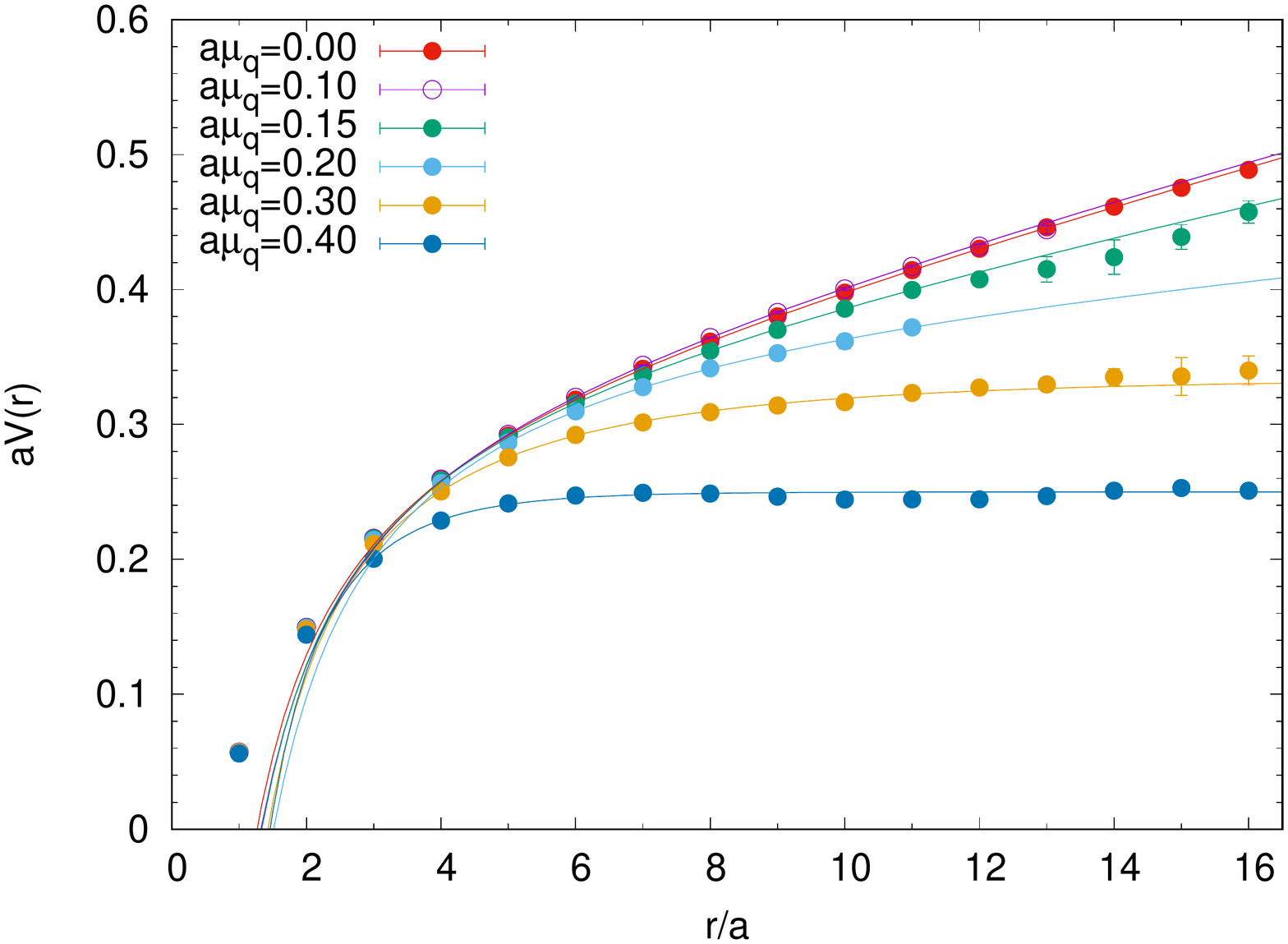}
\caption{The static potential $aV(r)$ as function of distance $r$ for few values of $\mu_q$
on $40^4$ lattices (left) and on $32^4$ lattices (right). The curves show
fits (\ref{eq1}) or (\ref{eq2}) as described in the text.}
\label{figure1}
\end{center}
\end{figure}

The static potential $V(r)$  is shown for few values of $\mu_q$ in the Fig.~\ref{figure1}(left)
for lattice $40^4$ and in the Fig.~\ref{figure1}(right) for lattice $32^4$. The curves show
results of the fit to the function
\beq
V(r) = V_0 + \sigma r + \alpha/r
\label{eq1}
\eeq
for $a\mu_q \le 0.35$ ($40^4$ lattices) or for $a\mu_q \le 0.25$ ($32^4$ lattices)  or to function
\beq
V(r) = V_0 + \alpha \frac{e^{-Er}}{r}
\label{eq2}
\eeq
for higher $a\mu_q$ values.

\begin{figure}[tbh]
\begin{center}
\includegraphics[width=10cm, angle=270]{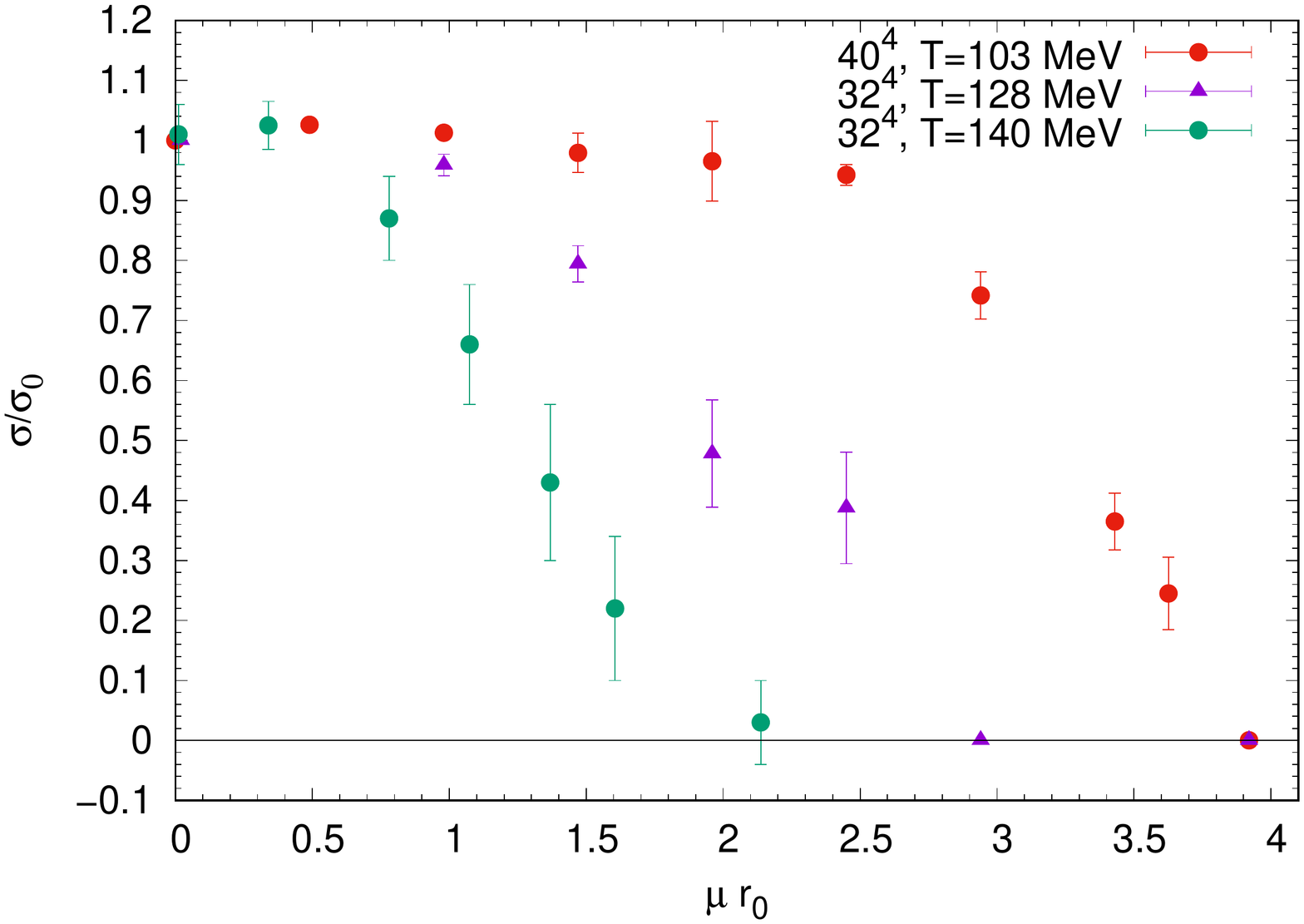}
\caption{The string tension $\sigma$ (devided by respective value obtained at $\mu_q=0$) as function of $\mu_q$ at three values of temperature. The results for $T=140$~MeV are taken from \cite{Bornyakov:2017txe}.}
\label{figure2}
\end{center}
\end{figure}
 
 We polt the string tension $\sigma$ dependence on $\mu_q$ for these two lattices in 
the Fig.~\ref{figure2}. Additionally we show the result from \cite{Bornyakov:2017txe}.
Note that for all three sets of data $\sigma$ is normalized by respective values $\sigma_0$ at $\mu_q=0$. The values of  $a^2\sigma_0$ computed on $40^4$ and $32^4$ lattices  at the same parameters are equal within error bars. One can see that for all three values of temperature there is a range of $\mu_q$ values where the string tension is not changing. Then it starts to decrease and goes to zero at some value of $\mu_q$ which can be defined as a confinement-deconfinement transition point.
As it was found in \cite{Bornyakov:2017txe} above this transition the static potential
can be described by the screened potential with screening mass increasing with increasing $\mu_q$. 

\begin{figure}[tbh]
\begin{center}
\includegraphics[width=6.5cm, angle=270]{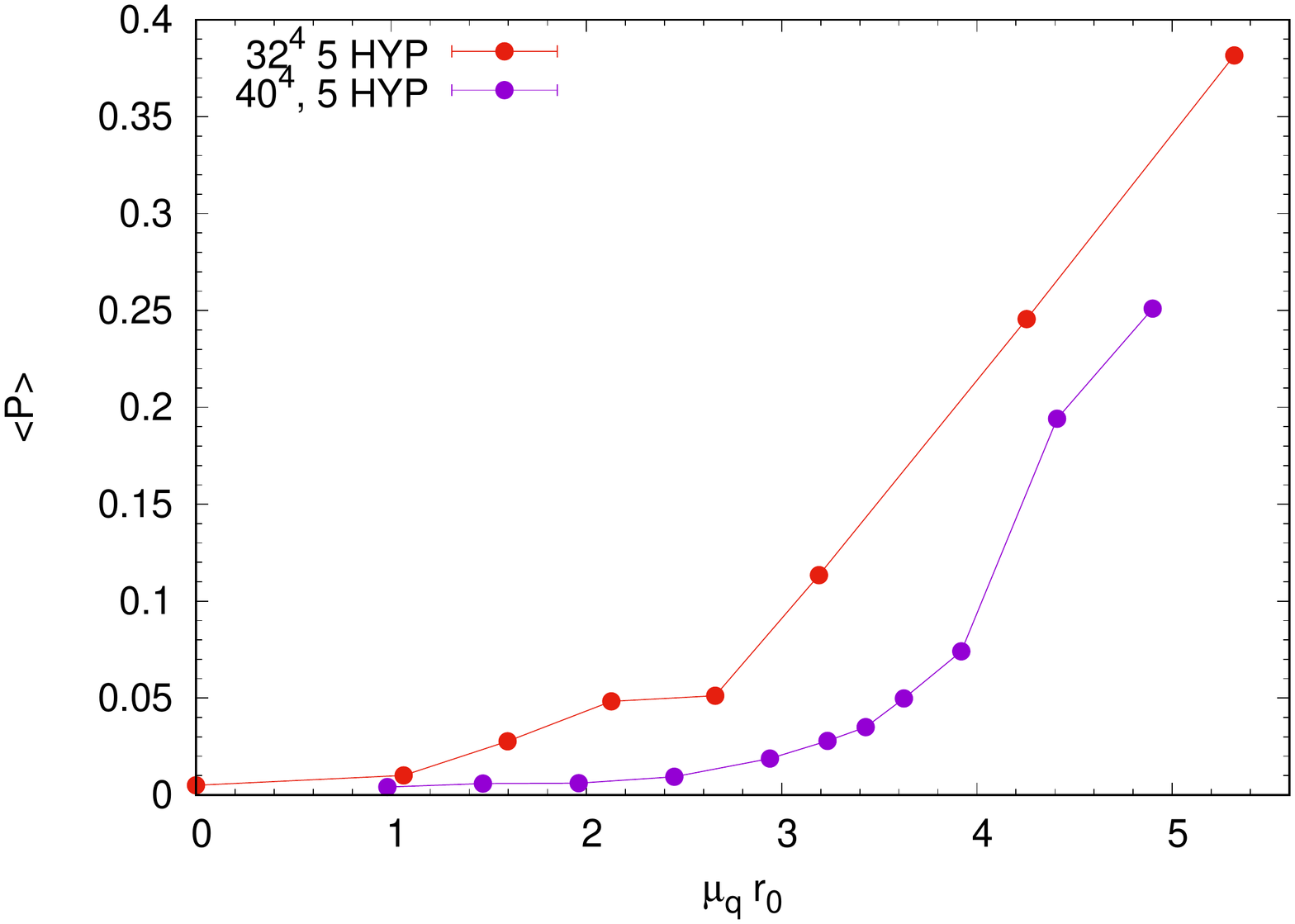} \hspace{-1.7cm}
\includegraphics[width=6.5cm, angle=270]{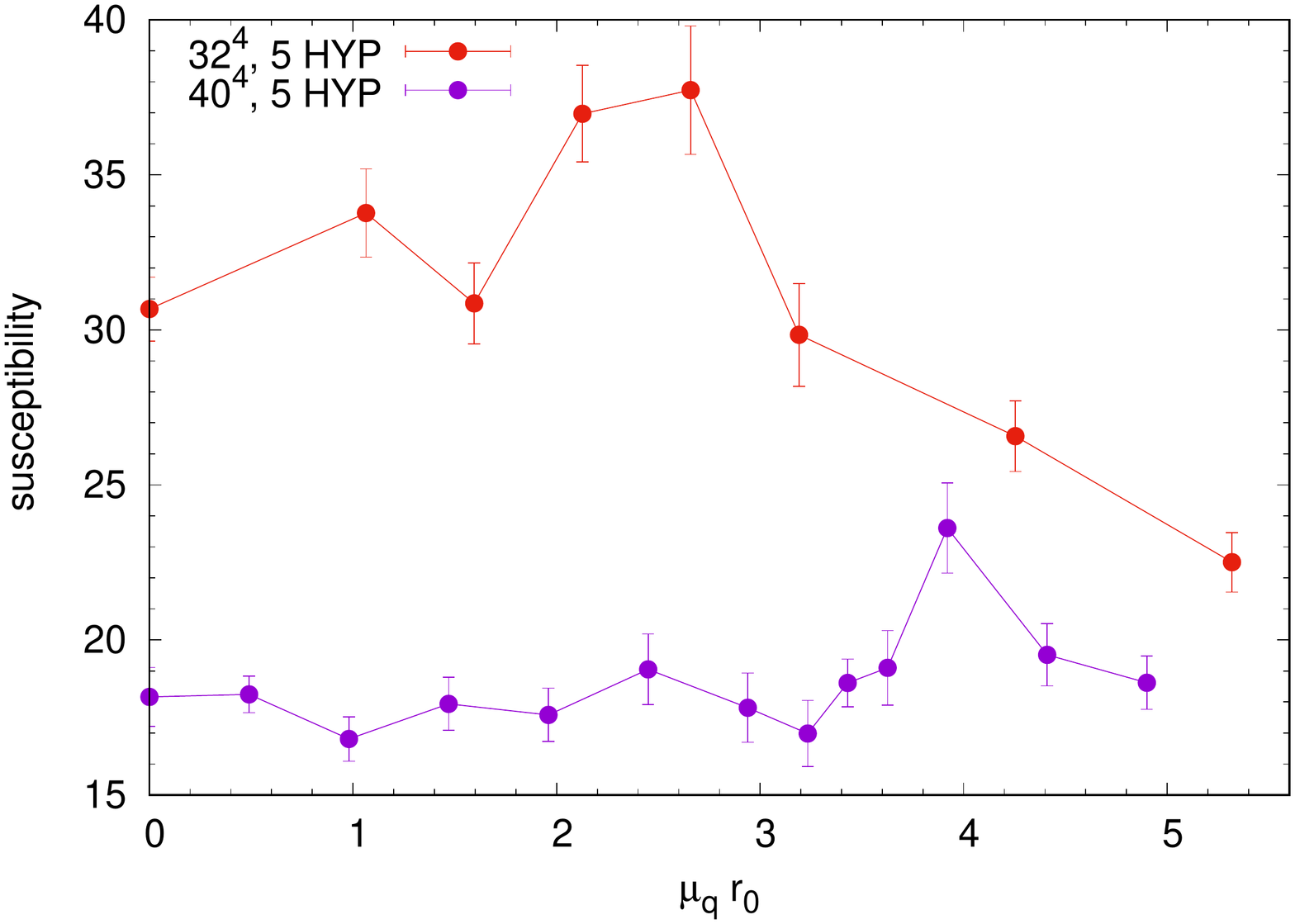}
\caption{The Polyakov loop (left) and its susceptibility (right) for two lattices computed after 5 HYP iterations.}
\label{figure3}
\end{center}
\end{figure}

Another way to determine the confinement-deconfinement transition is to use the Polyakov loop and its susceptibility. The Polyakov loop was used in particular in Ref.~\cite{Cotter:2012mb}. The Polyakov loop $P$ is defined as ($N_s=L/a)$
\beq
P = \frac{1}{N_s^3} \sum_{\vec{x}} \frac{1}{2} \text{Tr} \prod_{t=1}^{N_t} U_4(\vec{x},t) 
\eeq
To measure the average Polyakov loop $<P>$ and its susceptibility $\chi$ defined as
\beq
\chi = N_s^3 (<P^2> - <P>^2)
\eeq
we used from zero up to five HYP iterations. 
 We found that without HYP it is not possible to draw any conclusions about 
dependence of $<P>$ and $\chi$ on $\mu_q$ because of large statistical errors. 
One iteration of HYP did not help much. Starting from two HYP iterations we observed results which are qualitatively similar to those presented in the  Fig.~\ref{figure3}  where we show
our results for five HYP iterations. We found that 
the relative statistical errors for both $<P>$ and $\chi$
are slowly decreasing with increasing of the number of HYP iterations. 
We decided to stop at five HYP iterations since we did not expect substantial improvement after further increasing of the number of HYP iterations.
One can see from Fig.~\ref{figure3}(left)  that the rising of $<P>$ starts earlier for lattice $32^4$ than for $40^4$. 
This is consistent with the behavior of susceptibility $\chi$ depicted in Fig.~\ref{figure3}(right).
The positions of maxima of $\chi$ for both $32^4$ lattice 
and $40^4$ lattice are in a good qualitative agreement with values of $\mu_q r_0 $ where
the string tension turns zero. Thus the Polyakov loop indicates the confinement-deconfinement transition at about same values $\mu_q$ and these values are definitely temperature dependent. 

\begin{figure}[tbh]
\begin{center}
\includegraphics[width=10cm, angle=270]{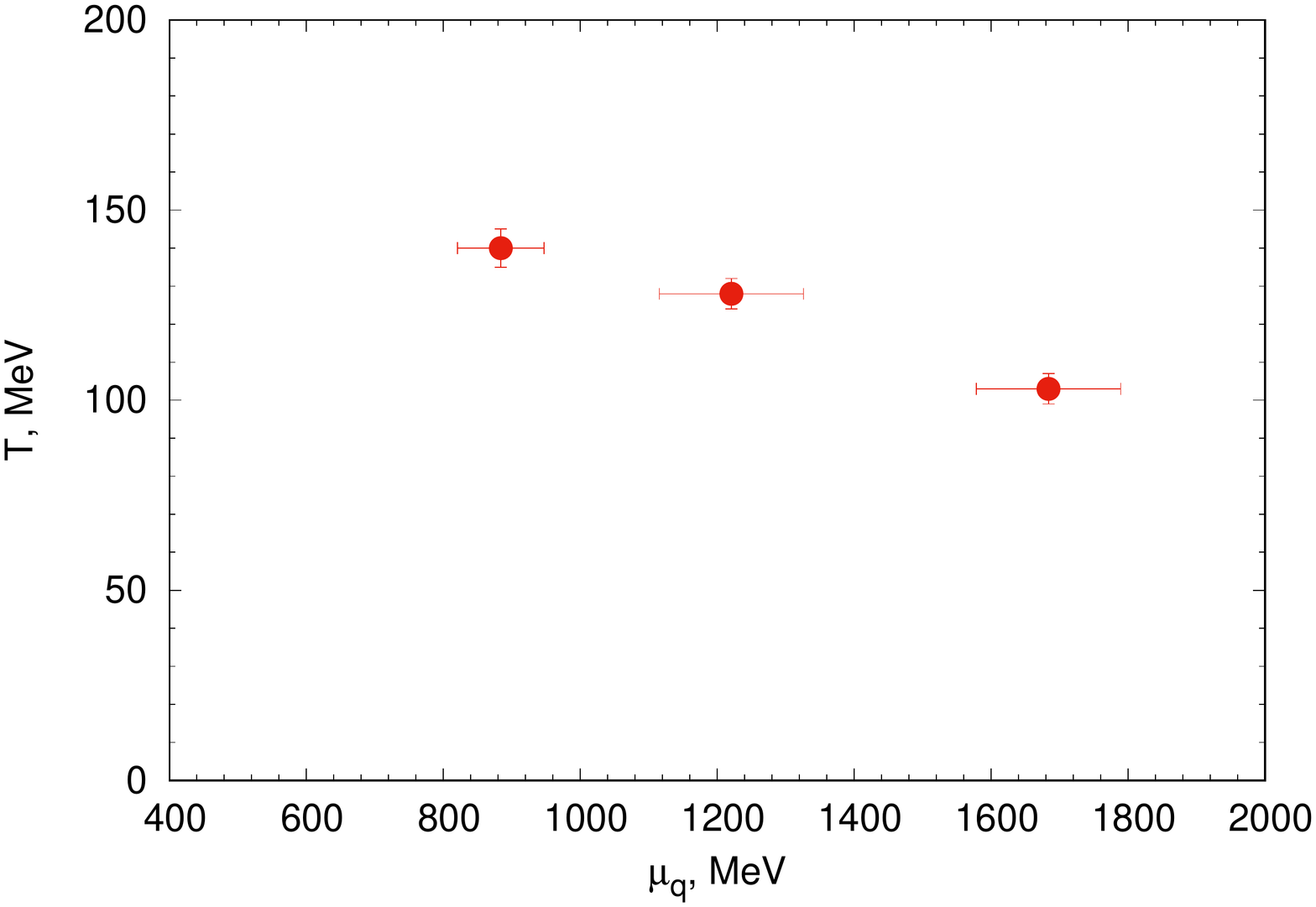}
\caption{The confinement-deconfinement transition in $(\mu_q, T)$ plane.}
\label{figure4}
\end{center}
\end{figure}

In Fig.~\ref{figure4} we  present our results for the confinement-deconfinement transition line in the $(\mu_q,T)$ plane. We take as a central value along the $\mu_q$ axes the minimal $\mu_q$ where the string tension turns zero and the error bars for this axes are defined by the distance to the nearest data point. These error bars cover the difference in the transition values determined from the string tension and from the Polyakov loop susceptibility. The data point from Ref.~\cite{Bornyakov:2017txe} is also used. The transition line is in a qualitative agreement with result obtained in 
Ref.~\cite{Cotter:2012mb}. But our result is shifted to higher $\mu_q$ values by factor two, approximately. We believe this quantitative difference is due to use of rather large lattice spacing in Ref.~\cite{Cotter:2012mb}. The fit with quadratic dependence 
on $\mu_q$ predicts zero temperature  confinement-deconfinement transition at $\mu_q$ value near to  2.5~Gev. Further studies on lattices with smaller temperature are needed to check and improve  this prediction.

\section{Conclusions}
Thus we observed that the confinement-deconfinement transition in the low temperature QC$_2$D is moving to higher values of $\mu_q$ when the temperature is decreasing. 
This phenomenon was demonstrated with the use of three observables: the string tension computed from the Wilson loops, the Polyakov loop and its susceptibility.
Our result is in a qualitative agreement with the earlier result of Ref.~\cite{Cotter:2012mb}.
But quantitatively our result for $\mu_q$ value at transition differs quite substantially, by factor two approximately.
We believe that the reason of this difference is that we used much smaller (by factor 4 approximately) lattice spacing. It is interesting to check the temperature dependence in this low temperature range for other important quantities, e.g. for the equation of state. This will be presented in a forthcoming paper.

  \acknowledgments{The authors are grateful to V. Braguta and A. Nikolaev for useful discussions.
This work was completed due to support of the Russian Foundation for Basic Research via grant 18-02-40130 mega
and is supported by Grant No. 0657-2020-0015 of the
Ministry of Science and Higher Education of Russia.
Computer simulations were performed at the Supercomputer SQUID (Osaka University, Japan), the FEFU GPU cluster Vostok-1, the Central Linux Cluster of the 
NRC ”Kurchatov Institute” - IHEP, the Linux Cluster of the NRC ”Kurchatov Institute” - ITEP (Moscow). In addition, we used computer resources of the federal collective usage center Complex for Simulation and Data Processing for Mega-science Facilities at NRC Kurchatov Institute, http://ckp.nrcki.ru/.}

\bibliographystyle{plain}
\bibliography{citations_asym_2016}

\appendix

\section{}
\label{sec_App}
Net quark density of noninteracting gas of massless quarks
has the form
\beq\label{eq:Omega_free}
n_q =  {g_f \over 6\pi^2}
\left( \mu_q^3 + \pi^2 T^2 \mu_q \right) \;,
\eeq
where $g_f=N_{spin}\cdot N_c \cdot N_{flavor}=8$ is the degeneracy factor
and for a rough estimate at low temperatures the second term can be neglected
(the more so taking it into account would only strengthen our conclusions);
quarks of mass 30~MeV at $\mu_q\sim 1$~GeV can be cosidered as massless.

Given $\mu_q=1.4$~GeV and $L=1.92$~fm, we arrive at 
$\ds n_q \sim 6 g_f\ \mbox{ fm}^{-3}$,
that is, all states corresponding to 42 lowest momenta 
$\ds \vec p={2\pi\over L} \vec n$ are occupied, 
where $\vec n = (n_1,n_2,n_3)$ runs over integer-valued 3D lattice.
Thus the Fermi surface embraces the sites 
$\vec n =(0,0,0), (0,0,1), (0,1,1), (1,1,1), (0,0,2)$ as well as 
those obtained from them by permutations of $n_1,n_2,n_3$ and/or
transformations $n_i\to -n_i$. Only part of momenta
corresponding to  $(n_1,n_2,n_3)=(0,1,2)$ is embraced by the Fermi surface. 
Excitation of lowest nonzero energy in this case
corresponds to the transition of the type $(0,1,2) \to (1,1,2)$, 
its energy is approximately $\ds E_{min}\simeq {p_{min}\over 10}\approx 60$~MeV,
which is even lower than
the temperature $\ds T={1\over 40 a}\approx {4.11\;\mbox{GeV}\over 40}\approx 103$~MeV.
Therefore, the temperature 
on the lattice under consideration cannot be considered as zero at 
large quark chemical potentials $\mu_q \stackrel{>}{\sim} 1$~GeV.
It should also be mentioned that the transitions of the type $(0,1,2) \to (1,1,2)$ and the like 
are rather numerous (several hundred for the $E_{min}$ only), what enhances the probability of such excitations.

\end{document}